\documentclass[fleqn,usenatbib,useAMS]{mnras}

\usepackage[T1]{fontenc}
\usepackage{ae,aecompl}

\usepackage{graphicx}	% Including figure files
\usepackage{amsmath}	% Advanced maths commands
\usepackage{amssymb}	% Extra maths symbols
\usepackage{natbib}

\title[A gas accretion scenario for planets in FARGO3D]{Adapting a gas accretion scenario for migrating planets in FARGO3D}
\author[L. A. DePaula et al.]{
L. A. DePaula,$^{1}$\thanks{E-mail: luiz.paula@usp.br}
T. A. Michtchenko,$^{1}$
\\
$^{1}$Instituto Astron\^omico, Geof\'isico e Ci\^encia astmosf\'ericas, Universidade de S\~ao Paulo, Rua do Mat\~ao 1226, 05508-900 S\~ao Paulo, Brazil\\
}

\date{Accepted 06/11/2018. Received 03/11/2018; in original form 08/08/2018}

\pubyear{2018}

\begin{document}
\label{firstpage}
\pagerange{\pageref{firstpage}--\pageref{lastpage}}
\maketitle

\begin{abstract}
%FARGO3D has been proposed to investigate numerically the gravitational interaction of the planet with the gas disc, 
%providing original outcomes. However, FARGO3D does not consider the gas accretion of the planet that may affects 
%the planetary migration process.

%The aim of this paper is to include a model of the gas accretion in the FARGO3D code. For this task, we choose Russell's scheme \citep{Russell2011}, which is an adaptation of Kley's model \citep{Kley1999} for the gas accretion onto migrating planets.
%With this, we are able to investigate the gas accretion process for a wide range of planetary masses.

%Initially, we study the influence of the gas accretion onto the planet on type II migration of giant planets, comparing our results with those obtained in \cite{Durman2015}. For this purpose, we follow the evolution of the planets in a two-dimensional locally isothermal disc with a specific accretion rate and different values of viscosity and planetary mass. Then, we extend our study to estimate the time needed for a low mass planet to open a gap in the gas disc, and compare its value with the characteristic time of type I migration.
FARGO3D has been proposed to investigate numerically the gravitational interaction of the planet with the gas disc, 
providing original outcomes. However, FARGO3D does not consider the gas accretion of the planet that may affect 
the planetary migration process. Thus, the aim of this paper is to include a model for the gas accretion in the 
FARGO3D code. For this task, we choose Russell's scheme, which is an adaptation of Kley's model for the gas 
accretion onto migrating planets. Initially, we study the influence of the gas accretion 
onto the planet on type II migration of giant planets.
For this purpose, we follow the evolution of the planets in a two-dimensional locally isothermal 
disc with a specific accretion rate and different values of viscosity and planetary mass considering 
two scenarios. In the first, the gas mass is withdrawn from the disc and is not added to the planet. In the 
second the planets migrate, while their masses grow due to the gas accretion.
Then, we extend our  study to estimate the time needed for a low mass planet to open a gap in the gas disc, and compare its 
value with the characteristic time of type I migration.
\end{abstract}

\begin{keywords}
Protoplanetary discs -- Planet disc interaction -- Planet formation
\end{keywords}

%%%%%%%%%%%%%%%%%%%%%%%%%%%%%%%%%%%%%%%%%%%%%%%%%%

%%%%%%%%%%%%%%%%% BODY OF PAPER %%%%%%%%%%%%%%%%%%

\section{Introduction}
   According to the classical theory of planetary formation, giant planets form in outer regions of the protoplanetary disc where
   there is enough material available \citep{Lissauer1993,Pollack1996}.    However,  a large number of giant planets with very short period orbits,    known as ``hot Jupiters'', has been discovered \citep[e.g.][]{Udry2007}. One of the mechanisms,  proposed to explain the locations   of such planets, is the planetary migration resulting from the gravitational interaction of the planet with the    gas disc \citep{Masset2000A,Benitez2011,Kley2012,Baruteau2013a,Baruteau2013b}.    Indeed, this interaction leads to an exchange of the angular momentum between the disc and the planets, changing their orbital parameters and resulting, in general, in the reduction of the planet's semiaxes and damped eccentricities \citep{Goldreich1980,Ward1986,Meyer1987,Lin1993,Ward1997}.

   Several numerical codes have been employed to investigate the gravitational interaction of the    planet with the gas disc, such as NIRVANA \citep{Ziegler1998} and FARGO \citep{Masset2000A}, among others.    However, the majority of the codes does not consider gas accretion onto the planet or deal with    simplified models. This is due to the fact that    the accretion process depends on several aspects of the disc structure and on a variety of physical    processes and their consideration  increases significantly computational costs.    Therefore, little is known about how the accretion process affects the local structure around the    planet, and, consequently, how it affects the planetary migration.
   Moreover, according to the classical theory of planetary formation, the rate of the gas accretion onto    the planet also depends on the mass of the planet. In fact, for low mass planets, the gas accretion process    is regulated by the Kelvin-Helmholtz contraction of the planet's gas envelope. On the other hand, as    the mass increases, the hydrostatic equilibrium breaks up and, as a consequence, the planet enters in a runaway
   accretion regime, at which the planet's mass increases very fast.

There are two main migration regimes known as type I and II migration.    Type I migration occurs for low-mass planets ($\lesssim$ 50$\mathrm{M}_{\mathrm{Earth}}$),    whose interaction with the disc is weak enough to leave the disc structure almost unperturbed. This type of migration can be approximated analytically by the linear regime \citep{Meyer1987}, when the total torque    on the planet due to the gas is defined by the combined effects of the Lindblad and corotation resonances \citep{Ogilvie2006}.    \cite{Goldreich1980} were the first to formulate
   the disc-planet interaction and derive formulae for the migration process. Later, \cite{Tanaka2002}
   extended the linear regime of excitation to three-dimensional isothermal discs and recalculated    the torques of the Lindblad and corotation resonances.

   On the other hand, type II migration occurs for high-mass planets (> 0.5 $\mathrm{M_{J}}$).    In this regime, a planet opens an annular gap, within which the disc surface density  is reduced  substantially from its unperturbed value; as a consequence, the gas in this region
   contributes less to the torque on the planet, thus reducing the amount of total torque of the system \citep{Bryden1999,Ida2004a}.
   The gap is responsible for a coupling between the orbital migration of the planet and
   the viscous evolution of the protoplanetary disc. So, in the case when the gas does not cross the gap,
   the planet migrates inward in a timescale determined by the disc's viscosity \citep{Lin1986,Ward1997}.
   However, studies show that the gas can cross the gap via horseshoe orbits, implying that the rate of migrates are different than those of the viscous evolution of the disc \citep{Edgar2004, Durman2015}.

   The calculation of the migration rate of the planets of intermediate mass is complicated because this process cannot be fitted by neither the  type I nor the type II migration, thus this issue still remain open \citep{Armitage2010}.

   Numerical simulations allow us to investigate how the accretion of gas onto the planet affects the    flow of gas in the gap, modifying the planet's migration rate. For example, \cite{Nelson2000},    using the method originally described by \cite{Kley1999} for gas accretion of the planet, have found that the orbital migration of massive planets is always inwards and that the planet reaches the central star after $10^{4}$ initial orbital
   periods, regardless of the fact that the planet is accreting mass or not.    Later, \cite{Durmann2017}, adding some refinements for the gas accretion to the same model, have   shown that a fast migrating planet is able to accrete more gas than a slowly migrating planet.
   Moreover, they showed that the amount of gas crossing the gap changes during the migration process    as the migration slows down.

   This work investigates type II migration considering the model for gas accretion presented in    \cite{Russell2011} and adapting it to suit the FARGO3D code \citep{Benitez2015}. Indeed, the FARGO3D code does not include a model for gas accretion, although, this process can affect the planetary migration.   Following \cite{Russell2011}, we use a timescale of gas accretion that considers    the orbital period of the gas around the planet. Additionally, we 	 using    the minimum value between the Bondi radius and the
   Hill radius. Finally, we introduce a gas accretion rate to deal with small planets, using the Kelvin-Helmholtz contraction time scale for planet's gas
   envelope. In particular, we investigate the influence of Russell's gas accretion model on the type II migration,    and the time for a low mass planet to open a significant gap in the disc.

   This manuscript is organized as follows. In Sect. 2, we describe the gas accretion model and introduce  the numerical model for the disc in Sect. 3.    In Sect. 4, we discuss how the gas accretion model affects the migration of giant planets and compare    our results with those reported in \cite{Durman2015} and \cite{Durmann2017}. In Sect. 5, we investigate the time it takes    for a accreting planet with initial mass $\simeq 20\mathrm{M_{Earth}}$, on a fixed orbit, to open a gap in the gas disc. Finally, we discuss the results in Sect. 6.

\section{Gas accretion scenario}\label{modelaccretion}

\cite{Kley1999} was one of the first works to introduce a model for gas accretion to a planet forming in the disc.  \cite{Russell2011}  modified the model including Bondi's radius for gas accretion zone, effects of the disc height, Kelvin-Helmholtz gravitational contraction time scale, among others.
The calculation of the Bondi's radius is  a condition for the calculation of the gas accretion zone.
Bondi's radius determines the region around the massive body immersed in the gas cloud, from which the body captures the gas. By definition, in this region, the resulting (thermal + relative to the body) velocity of the gas  is smaller than the escape velocity of the gas \citep{Bondi1952}. This condition provides that

\begin{equation}
   R_{B_{ij}} = \frac{2GM_{\mathrm{p}}}{c_{\mathrm{s}}^{2} + \Delta v_{ij}^{2}},
\end{equation}

\noindent{where $\Delta v_{ij}$ is the module of the relative velocity between the planet and gas residing in cell $(i,j)$,
$c_{\mathrm{s}}$ is speed of sound in the cell, $G$ is the gravitational constant and $M_{\mathrm{p}}$ is the mass of planet.}

The radius of the accretion zone of  the planet, $R_{a_{ij}}$, is then determined by the condition

\begin{equation}\label{radiusaccretion}
   R_{a_{ij}} = \mathrm{min} (R_{B_{ij}}, R_{\mathrm{H}}),
\end{equation}

\noindent{where $ R_{\mathrm{H}}$ is the Hill radius of the planet given by}

\begin{equation}\label{hill}
    R_{\mathrm{H}} = a \left( \frac{M_{\mathrm{p}}}{3 M_{\star}} \right)^{1/2},
\end{equation}

\noindent{with $a$ and $M_{\star}$ denoting the semimajor axis of the planet and the mass of the star, respectively. Defined in this way, $R_{a_{ij}}$ determines a minimum distance at which a cell should be from the planet, to be considered within the accretion region of the planet. If the distance from the cell to the planet $r_{ij}$ is smaller than $R_{a_{ij}}$, then the gas will be removed from this cell during accretion.}

The mass of gas removed from a cell with position $[i,j]$ inside the accretion zone in a particular time step $dt$ (intrinsic time of the hydrodynamic code), is calculated as

\begin{equation}
    \delta m_{ij} = f_{ij} \Sigma_{ij} A_{ij} dt,
\end{equation}

\noindent{where $\Sigma_{ij}$ is the cell surface density, $A_{ij}$ is the cell surface area and $f_{ij}$ is the frequency
with which the gas is removed from the cell, defined by}

\begin{equation}\label{perT}
   f_{ij} = \frac{1}{2\pi} \sqrt{\frac{G M_{\mathrm{p}}}{r_{ij}^{3}}},
\end{equation}

\noindent{where $r_{ij}$ is the distance between the cell and the planet. Thus, the total mass, $\Delta m_{n}$, removed from all accreting cells per nth time-step , is a sum}

\begin{equation}\label{sumass}
    \Delta m_{n} = \sum_{i,j} \delta m_{ij},
\end{equation}
and the mass of planet is then increased by this amount.

The linear momentum is also transferred from the gas to the planet during accretion. However, according to results obtained by \cite{Durmann2017}, this process does not change significantly  the migration rate of the planet. So, in this paper, we choose to work only with the increase in planetary mass.

The application of the Bondi's radius to calculate the accretion zone results in a dynamic accretion rate, which varies around the planet. 
The model is adaptable for the runaway regime for gas accretion of giant planets. On the other hand, low mass planets are outside the runaway regime and, 
as shown in \cite{Russell2011}, do not accrete mass in a rate beyond the Kelvin-Helmholtz contraction time scale. Indeed, the Kelvin-Helmholtz gravitational 
contraction time scale, $\tau_{\mathrm{KH}}$, represents the characteristic growth time of a planet's gas envelope. \cite{Ikoma2000} found that 
$\tau_{\mathrm{KH}}$ is strongly dependent on the planetary mass and is expressed approximately as

\begin{equation}\label{KH}
   \tau_{\mathrm{KH}} \simeq 10^{b} \left( \frac{M_{\mathrm{p}}}{M_{\star}} \right)^{-c} \left( \frac{\kappa}{1 \mathrm{cm}^{2} \mathrm{g}^{-1}}\right) \mathrm{yr},
\end{equation}

\noindent{where $\kappa$ is the grain opacity with power index $b = 8$ and $c = 2.5$.
In the present work, however, we adopt an approach used in \cite{Ida2004a}, setting $\kappa = 1.0~\mathrm{cm}^{2} \mathrm{g}^{-1}$, which gives}

\begin{equation}\label{tau}
   \tau_{\mathrm{KH}} \simeq 10^{8} \left( \frac{M_{\mathrm{p}}}{M_{\star}} \right)^{-2.5} \mathrm{yr}.
\end{equation}
This expression gives an approximate growth time of the gas envelope for a planet with low mass.

According to \cite{Russell2011}, the rate which limits gas accretion onto the planet is determined by the minimum
between the potential gas accretion rate ($\dot{M}_{\mathrm{U}}$) and the Kelvin-Helmholtz gas accretion
rate ($\dot{M}_{\mathrm{KH}}$) for a particular time step $dt$ (intrinsic time of the hydrodynamic code):

\begin{equation}\label{limitingaccretion}
   \dot{M} = \mathrm{min} (\dot{M}_{\mathrm{U}}, \dot{M}_{\mathrm{KH}}) = \mathrm{min} \left(\frac{\Delta m_{n}}{dt},\frac{M_{\mathrm{p}}}{\tau_{\mathrm{KH}}}\right) ,
\end{equation}

\noindent{where $\tau_{\mathrm{KH}}$ is given by Eq. \ref{tau} and $\Delta m_{n}$ by the Eq. \ref{sumass}.}

The accretion rate given by Eq. \ref{limitingaccretion} determines the ratio of gas removed from the cells
within the planet's accretion zone. The ratio of gas accretion, $\mu$, is:

\begin{equation}
   \mu = \frac{\dot{M}_{\mathrm{KH}}}{\dot{M}_{\mathrm{U}}},
\end{equation}

\noindent{where $\mu \leqslant 1$. The mass removed from each cell in the planet's accretion zone is reduced
by $\mu$. Thus,}

\begin{equation}
   \delta m_{ij} = \mu f_{ij} \Sigma_{ij} A_{ij}.
\end{equation}

This equation implies that, if $\mu < 1$, $\dot{M}_{\mathrm{KH}} < \dot{M}_{\mathrm{U}}$, and,
consequently, the Kelvin-Helmholtz rate limits the gas accretion. Otherwise, $\mu = 1$ and the
accretion of gas occurs in the runaway regime.

In summary, unlike the scenario described in \cite{Kley1999}, the model described by Russel allows us to work with 
the accretion mechanism for small planetary masses using the Kelvin-Helmholtz contraction time scale for planet's gas envelope. 
Besides that, it is possible to adapt the size of the accretion zone taking into account the radius of Bondi. In addition, 
Russell's model uses, for the runaway regime, a timescale of gas accretion that considers the orbital period of the gas around 
the planet that results in a dynamic range of accretion time scales for each cell where gas is accreted.

In this work, Russel's model described above has been adapted in FARGO3D, for use in GPU programming.

\section{Numerical setup}

   To reduce high computational costs, we apply the hydrodynamic code FARGO3D \citep{Benitez2015} in the 2D regime in our simulations. The advantage of using FARGO3D in 2D setup, instead of using its predecessor FARGO, is that FARGO3D allows GPU computation, thus reducing computational time. The application of the three-dimensional model to gas accretion study is left for future works.

   In order to compare results, we choose setup parameters similar to those adopted for the NIRVANA code in \cite{Durman2015}.    Assuming a thin gas disc, with typical vertical height ($H$) much smaller than its radius, we introduce a 2D approximation    of the problem.    The disc is considered to be locally isothermal, with viscosity defined by the parameter $\alpha$ from \cite{Shakura1973}.    The star, with $M_{\star} = 1\mathrm{M}_{\odot}$, is located at the origin of a cylindrical coordinate system, while the disc extends from 1.56 to 15.6 au (0.3 to 3.0 in code units). 
   The resolution of the disc is fixed at 582 $\times$ 1346 cells equally spaced and no significant differences ($< 5\%$) in the final position of the planet and the migration    time have been found using higher resolutions. The planet of mass $M_{\mathrm{p}}$ is placed initially
   on a circular orbit at a distance of $\backsimeq$5.2 ua, that corresponds to $r = 1.0$ code units.

   Following \cite{Durmann2017}, we work with a steady accretion flow through the disc, that is, with the constant local    accretion rate $\dot{m}$. In this situation, the radial and angular components of the velocity of gas, $v_{\mathrm{r}}$ and $v_{\theta} $, respectively,   are given by \citep{Durman2015}:

\begin{equation}\label{eqvr2}
    v_{\mathrm{r}} = - \frac{3}{2} \alpha h^{2} r \Omega_{\mathrm{K}},\\
    v_{\theta} = \sqrt{1 - \frac{3h^{2}}{2}} r \Omega_{\mathrm{K}},
\end{equation}

\noindent{where $\Omega_{\mathrm{K}} = \sqrt{GM_{\star}/r^{3}}$ is the Keplerian orbital frequency of the gas at the position $r$ and the parameter $h$ corresponds to the relative disc thickness and has constant value $h = H/r = 0.05$. The surface density of the disc is given by
\citep{Durman2015}:}

\begin{equation}\label{eqSigma}
    \Sigma = \frac{\dot{m}}{3\pi \alpha h^{2} \sqrt(G M_{\star})} r^{-1/2} = \Sigma_{0} r^{-1/2},
\end{equation}

\noindent{where $\Sigma_{0}$ is the superficial density at $r$ = 1 code units. These equations can 
be obtained for stationary accretion disks with constant accretion rate.}

Table \ref{table1} shows the parameter space used in our simulations. Throughout the paper we will
refer to the disc defined by the of parameters ($\alpha$, $q$, $\dot{m}$) = (0.003, 0.001, $10^{-7}$ $\frac{\mathrm{M}_{\star}}{\mathrm{year}}$) as the standard disc, with $q = M_{\mathrm{p}}/M_{\star}$.
It is worth mentioning that in the FARGO3D code the initial profile of the surface 
density of the disc is obtained through the free parameters for the disk aspect ratio ($h$ = 0.05), 
the flare of the disc ($\gamma$ = 0) and the surface density in $r$ = 1 ($\Sigma_{0}$).
Table \ref{table2} displays the values of $\Sigma_{0}$ obtained for each parameter set
($\alpha$,$\dot{m}$) according to Eq. \ref{eqSigma}. The values are in $\mathrm{kg} \cdot {\mathrm{m}}^{-2}$.

\begin{table}
\caption{Space of parameters. The values of $\alpha$ and $q$ ($q = M_{\mathrm{p}}/M_{\star}$) are
dimensionless. We define a standard parameter set as ($\alpha$, $q$, $\dot{m}$) = (0.003, 0.001, $10^{-7}$ $\frac{\mathrm{M}_{\star}}{\mathrm{year}}$).}             % title of Table
\label{table1}      % is used to refer this table in the text
\centering                          % used for centering table
\begin{tabular}{c c c}        % centered columns (4 columns)
\hline\hline                 % inserts double horizontal lines
$\alpha$ & $q$ & $\dot{m}$ $\left( \frac{\mathrm{M}_{\star}}{\mathrm{year}} \right)$  \\    % table heading
\hline                        % inserts single horizontal line
   0.001 & 0.0002 & $1 \times 10^{-9}$\\      % inserting body of the table
   0.003 & 0.0005 & $1 \times 10^{-8}$\\
   0.01  & 0.001  & $1 \times 10^{-7}$\\
\hline                                   %inserts single line
\end{tabular}
\end{table}

\begin{table}
\caption{Values for $\Sigma_{0}$ in  $\mathrm{kg} \cdot {\mathrm{m}}^{-2}$ for each set of parameters.}             % title of Table
% is used to refer this table in the text
\label{table2}      % is used to refer this table in the text
\centering                          % used for centering table
\begin{tabular}{c | c c c}        % centered columns (4 columns)
\hline\hline                 % inserts double horizontal lines
  $\alpha$ $\textbackslash$ $\dot{m}$ & $1 \times 10^{-7}$ $\frac{\mathrm{M}_{\star}}{\mathrm{year}}$ & $1 \times 10^{-8}$ $\frac{\mathrm{M}_{\star}}{\mathrm{year}}$ & $1 \times 10^{-9}$ $\frac{\mathrm{M}_{\star}}{\mathrm{year}}$\\    % table heading
\hline                        % inserts single horizontal line
         &           &      & \\      % inserting body of the table
   0.001 &     2626.3  &    262.63     &    26.263   \\
   0.003 &     8743.4  &    874.34     &    87.434   \\
   0.010 &     2626.3  &    262.63     &    26.263   \\
\hline                                   %inserts single line
\end{tabular}
\end{table}

We modify the FARGO3D code in such a way that the torque on the planet due to the gas is reduced
when the gas is inside a region with radius equal to $0.8R_{\mathrm{H}}$ around the planet.
Following \cite{Durmann2017}, we use a Fermi-like tapering function the  given by

\begin{equation}\label{eqatenuacao}
    f_{\mathrm{taper}}(r_{\mathrm{cell}}) = \frac{1}{1 + \mathrm{exp}\left(-\frac{r_{\mathrm{cell} - 0.8R_{\mathrm{H}}}}{0.8R_{\mathrm{H}}}\right)},
\end{equation}

\noindent{where $r_{\mathrm{cell}}$ is the distance of the cell to the planet and $R_{\mathrm{H}}$ is the planet's Hill radius.}

For comparison purpose, we choose the same boundary conditions used by \cite{Durman2015}. So, we disable the damping of the surface density in the FARGO3D code, to avoid accumulation of gas at the inner boundary and allow gas to flow freely as desired.

\section{Giant planet migration}\label{giantplanets}

In order to investigate how Russell's accretion model \citep{Russell2011} affects the type II migration, we analyse two scenarios for the migration of giant planets. In the first scenario (Sec. \ref{withoutgroth}), the planets do not grow during migration, that is, the gas mass is withdrawn from the disc  and is not added to the planet. In the second scenario (Sec. \ref{withgroth}), the planets migrate, while their masses grow due the gas accretion.

First, we need to know how the gas accretion of the planet affects the profile of the gap, also as the flow of gas through the gap opened by the planet modifying the torque on the planet. Additionally, we analyse the rate of planetary migration for different values of the disc viscosity. In this
work we have three conditions for the viscosity of the disc, we use the term low viscosity when $\alpha = 0.001$, medium viscosity
when $\alpha = 0.003$ and high viscosity when $\alpha = 0.01$.

\begin{figure}
\centering
\includegraphics[width=\hsize]{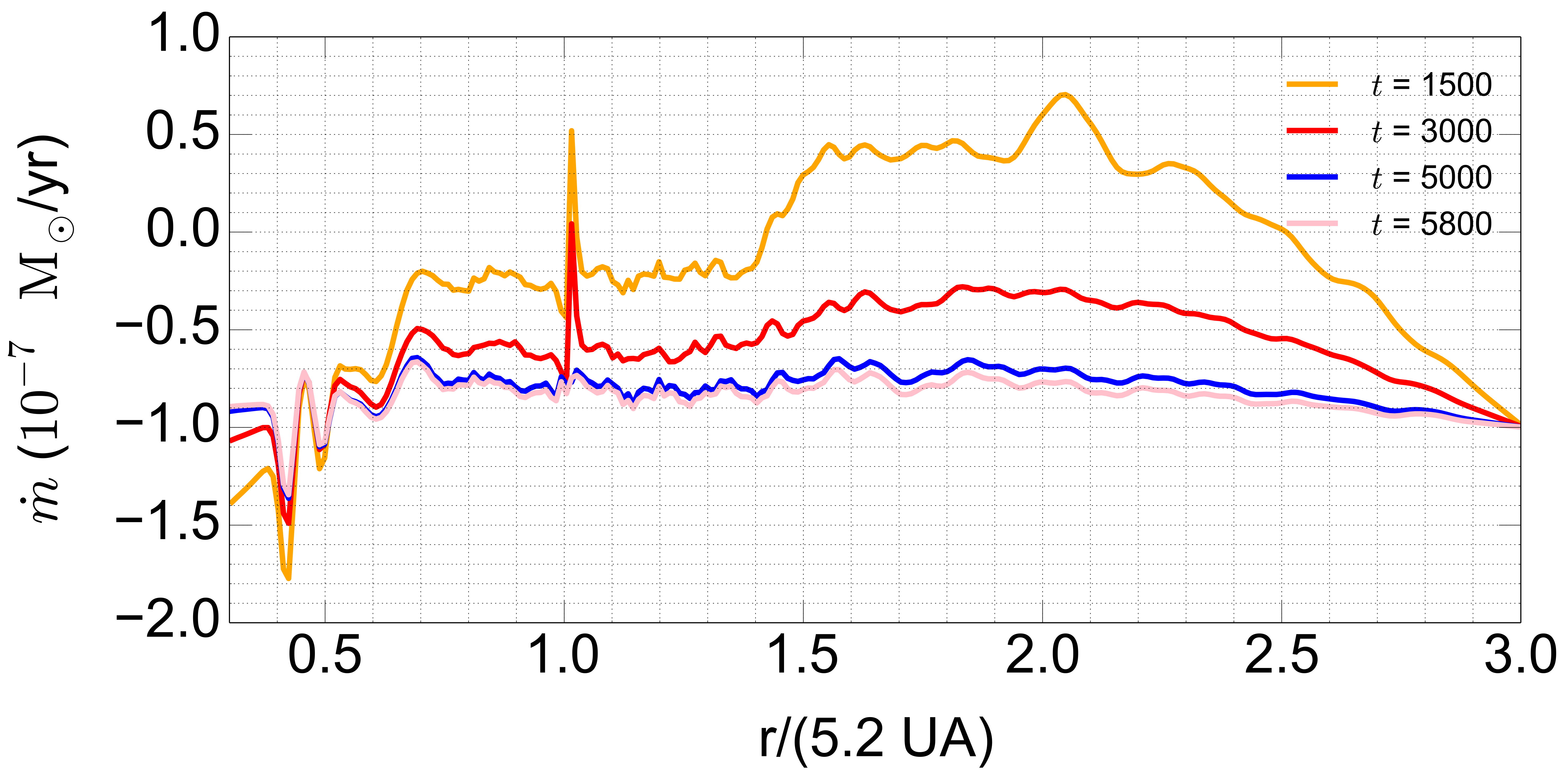}
\caption{Profiles of the local accretion rate obtained for the standard parameter set ($\alpha$, $q$, $\dot{m}$)
      = (0.003, 0.001, $10^{-7}$ $\frac{\mathrm{M}_{\star}}{\mathrm{year}}$), with a planet at $r$ = 1.
      The profiles are shown at four different instants  after the beginning of the calculation. Time is dimensionless and is defined in terms of the number of orbits of the planet fixed at $r$ = 1 code units.}
\label{fig1}
\end{figure}

These analyses require the gas disc to be in equilibrium. Thus, following the approach of \cite{Durman2015}, we calculate the time needed for the disc to reach the steady state. Figure~\ref{fig1} shows the profiles of $ \dot{m} (r)$ as a function of the radius of the disc after  different time periods, calculated for the planet of 1.0$\mathrm{M_{J}}$ fixed at $r = 1$ code units. To calculate the local accretion rate, we include a specific subroutine in the FARGO3D code, which  integrates the azimuthal mass fluxes produced by the code.

As seen in Figure~\ref{fig1}, although the profile of the accretion rate does not attain a constant
value, it converges to a nearly flat function between 5000 and 5800 orbital periods of the disc.
Thus, we can assume that the disc reaches equilibrium state after 5000 orbits and adopt this value
as a standard timespan.

\begin{figure*}
\resizebox{\hsize}{!}
{\includegraphics[width=\hsize]{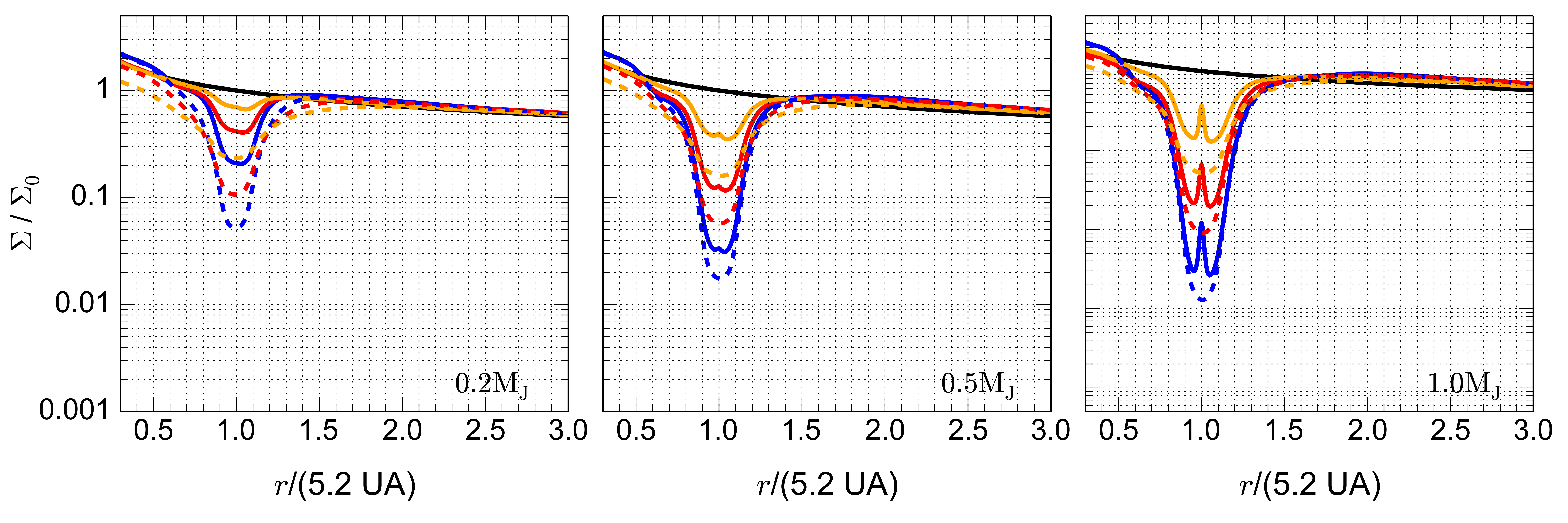}}
\caption{Azimuthally averaged surface density for non-accreting (solid line) and accreting (dashed line) planets
for different values of the planet mass (left: $\simeq 0.2\mathrm{M_{J}}$ or $q = 0.0002$, middle: $\simeq 0.5\mathrm{M_{J}}$
 or $q = 0.0005$ and right: $\simeq 1\mathrm{M_{J}}$ or $q = 0.001$). The blue lines corresponds  to $\alpha = 0.001$, the red lines to $\alpha = 0.003$ and orange lines $\alpha = 0.010$. The black line indicates the initial density at the beginning of the
simulation without planet.}
\label{fig2}
\end{figure*}

Figure~\ref{fig2} shows the azimuthally averaged surface density for non-accreting (solid lines) and accreting (dashed lines) planets, obtained after 5000 orbital periods of the disc,  for the three values of planetary masses and three different viscosity regimes used in our simulations. As expected, we observe the completely developed gaps in all cases. The gaps are increasing in depth with the increasing planet mass and the decreasing viscosity of the disc. The regions of overdensity at the positions of the planets ($r = 1$ code units) appear in the case when accretion processes are not considered (solid lines). This is due to the accumulation of gas around the non-accreting planet, treated as a point in the hydrodynamic mesh; when the gas accretion into the planet is taken into account, this artefact disappears (dashed lines).  Finally, we have performed the same simulations for different accretion rates of the disc, ($10^{-8}\mathrm{M}_{\star}\mathrm{yr}^{-1}$ and $10^{-9}\mathrm{M}_{\star}\mathrm{yr}^{-1}$), but we have not observed any significant difference; the same behaviour is reported for the NIRVANA code in \cite{Durman2015}.

\subsection{Migrating planets without growth}\label{withoutgroth}

Soon after the gas disc reaches equilibrium, the planets are allowed to migrate. We followed the migration
tracks of planets for the parameter sets shown in Table \ref{table1}. The measurements were performed for accreting and non-accreting planets during 1200 orbital periods of the disc; in the accreting case, the mass withdrawn from the disc was not added to the planet. %The results obtained  are shown in Fig. \ref{fig5}.

\begin{figure*}
\resizebox{\hsize}{!}
{\includegraphics[width=\hsize]{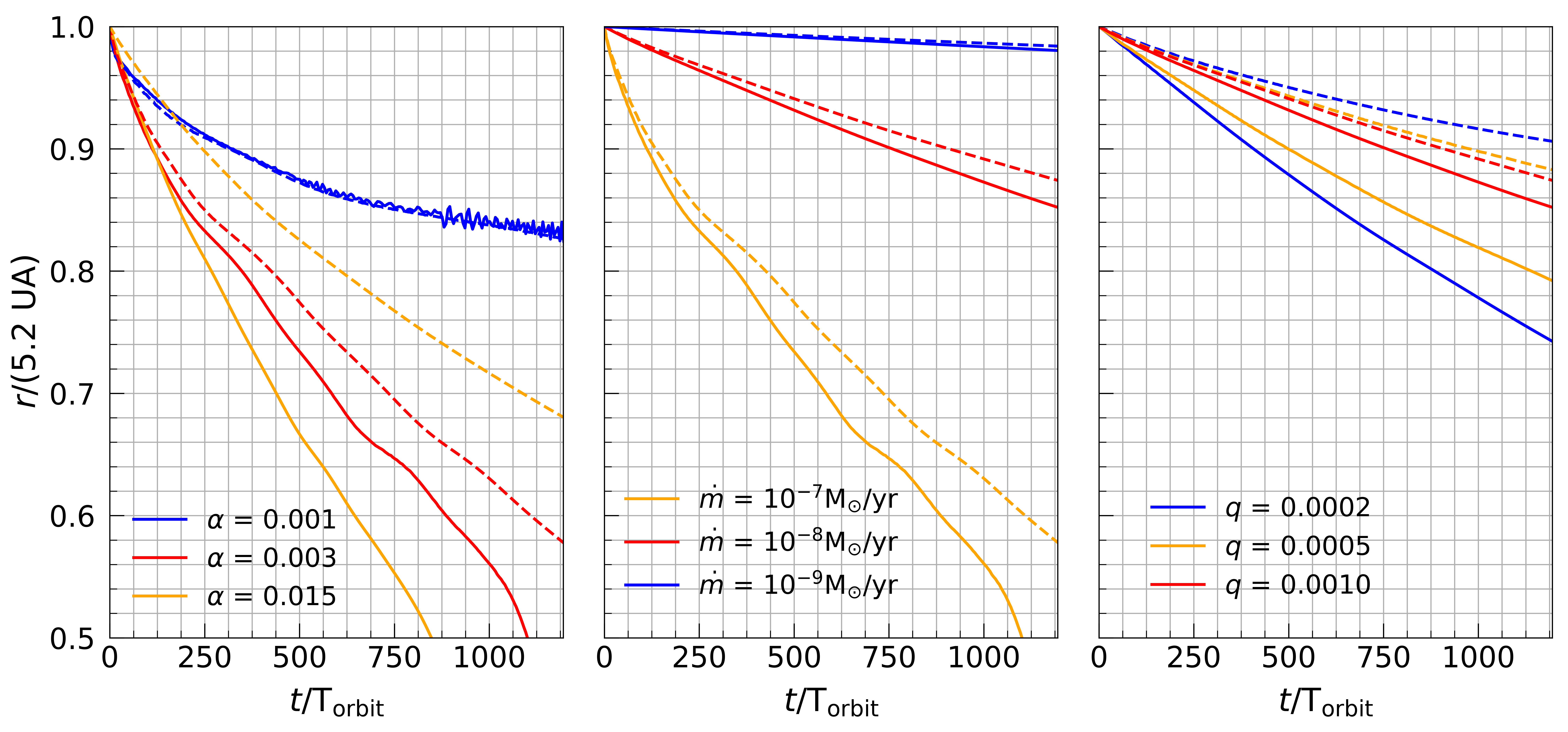}}
\caption{Planet migration tracks for different model parameters. In the left frame we have a planet mass of
$\simeq 1\mathrm{M_{J}}$ (or $q$ = 0.001) on a disc with accretion rate $10^{-7} \mathrm{M_{\odot}} / \mathrm {yr} $
for different viscosity values (blue: $\alpha = 0.001$, red: $\alpha = 0.003$ and orange: $\alpha = 0.015$).
In the central frame we have a planet mass of $\simeq 1\mathrm{M_{J}}$ (or $q$ = 0.001) on a disc with
$\alpha = 0.003$ and different accretion rates (orange: $10^{-7} \mathrm{M_{\odot}} / \mathrm {yr}$,
red: $10^{-8} \mathrm{M_{\odot}} / \mathrm {yr}$ and blue: $10^{-9} \mathrm{M_{\odot}} / \mathrm {yr}$).
In the right frame we have different planetary masses (blue: $\simeq 0.2\mathrm{M_{J}}$ or $q$ = 0.0002, orange:
$\simeq 0.5\mathrm{M_{J}}$ or $q$ = 0.0005
and red: $\simeq 1.0\mathrm{M_{J}}$ or $q$ = 0.001) in a disc with $\alpha = 0.003$ and accretion rate
$10^{-8} \mathrm{M_{\odot}} / \mathrm {yr}$.
The solid line is for non-accreting planets, and the dashed line for accreting planets.}
\label{fig5}
\end{figure*}

The migration tracks obtained are shown Fig. \ref{fig5}, where the parameters of the standard set are changed one-by-one, in order to analyse the behaviour of the planet.  We show the dependence of the rate of migration on the viscosity of the disc in the left frame of Fig. \ref{fig5}. In the case of high viscosity ($\alpha = 0.015$), there is a notable difference between the behaviour of accreting (dashed line) and non-accreting (solid line) planets. With decreasing $\alpha$, the difference also decreases and we note no difference in the case of low viscosity ($\alpha = 0.001$). %, for non-accreting planets the small oscillations disappear, showing that they are associated with the presence of mass in the region near the planet.

The middle frame of Fig. \ref{fig5} shows that the behaviour of the planet for different accretion rate of disc is only slightly different for accreting (dashed line) and non-accreting (solid lines) planets.

According to the right frame of Fig. \ref{fig5}, the effect of accretion process is much more significant for low-mass planets.  This is due to the fact that it accelerates the opening of the gap in a region where the gravitational effect of the planet is less intense.

As shown in Fig. \ref{fig5}, in general, the orbital decay of the planet is slower for accreting planets than for non-accreting planets. This difference is less significant with the increase of the planetary
mass (right frame) and for the decreasing accretion rate of disc (middle frame). This is due to the fact that the density along the disc decreases with the decrease of the rate of accretion (see Eq. \ref{eqSigma}); as a consequence, the total mass of the disc also decreases. Moreover, the gas mass in the nearby region of the planet also decreases when planets are accreting. In fact, this decrease in the gas mass near the planet causes the effect of the torque on the planet to decrease, thus, large planets in low mass discs have a smaller effect on their migratory process due to gas accretion. Similar results were obtained by \cite{Durman2015} using a different model for gas accretion.

It should be noted that some simulations were excluded from our analysis. This is the case when the planets escape from the gap, what occurs for the non-accreting planets with the masses of $0.2 \mathrm{M_{J}}$ and $0.5 \mathrm{M_{J}}$, in the low ($\alpha = 0.001$) and medium ($\alpha = 0.003$) viscosity regime. This also occurs for the planet with the mass of $0.2 \mathrm{M_{J}}$ for accreting planets in all viscosity regimes considered.

In what follows, we analyse how the gas accretion model influences both the gap profile and the mass fluxes in the gap.

\subsubsection{Gap profiles for accreting and non-accreting planets}

The gap created by the planet in the disc has a significant impact on the migration process because the decreasing superficial density of gas in the gap region reduces the contribution to the torque on the planet \citep{Durman2015}.

\begin{figure}
\centering
\includegraphics[width=1.0 \columnwidth,angle=0]{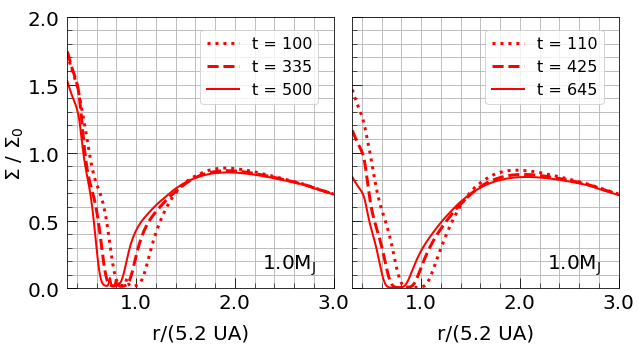}
\caption{Global gap profiles obtained for the standard parameter set, at three different times during planet migration.
The left frame corresponds to non-accreting planets, while the right frame corresponds to accreting planets.}
\label{fig6}
\end{figure}

Figure \ref{fig6} shows the time evolution of the gaps created by non-accreting planets (left panel) and  accreting planets (right panel). We compare the plots when the planet is in the same radial position, consequently, time is different in both plots. Looking at the planet at given positions helps to better understand the difference in local physical processes. 
For accreting planets, we note a larger lost of mass at the inner border of the disc due to the boundary conditions. When the planet is far away from the inner border of the disc, this effect does not influence on the profile of the gap.
A similar analysis was done for different parameter sets from Table 1 and the conclusion made is that effect is stronger in the higher viscosity regime and for lower accretion rate of disc.

\begin{figure}
\centering
\includegraphics[width=1.0 \columnwidth,angle=0]{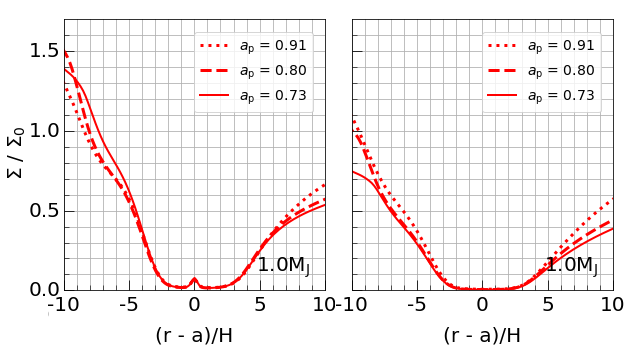}
\caption{Gap profiles obtained for the standard parameter set, at different positions of the planet in the disc during its migration path. The gap is re-scaled using the local disc scale height, $H$, at the position of the planet. The curves correspond directly to those shown in Fig. \ref{fig6}. The left frame corresponds to non-accreting planets, while the right frame corresponds to accreting planets.}
\label{fig7}
\end{figure}

Figure \ref{fig7} shows the local disc structure in the gap region obtained for the standard parameter set. The gap profile suffers no significant changes during the migration process of accreting and non-accreting planets.
In addition, in Fig. \ref{fig7}, we observe that the density bump (mass concentration around the position of the planet) is eliminated in the case of accreting planets. This feature remains during planetary migration and the same effect can be observed  for different values of the planetary mass and accretion rate of the disc.

Considering different values of the disc mass (i.e., accretion rate of the disc), we have found that, for fixed $\alpha$-value, the gap structure near the planet does not change considerably, in both cases of accreting and non-accreting planets. These results agree with the results obtained with the NIRVANA code in \cite{Durman2015}, where only non-accreting planets were investigated.

In summary, for accreting planets, our results show that, despite the accretion of gas generating deeper gaps,
the gas accretion does not alter significantly the gap profile around the migrating planet provided that
the gap is established.

\subsubsection{Flow across the gap for accreting and non-accreting planets}

\begin{figure}
\centering
\includegraphics[width=1.0 \columnwidth,angle=0]{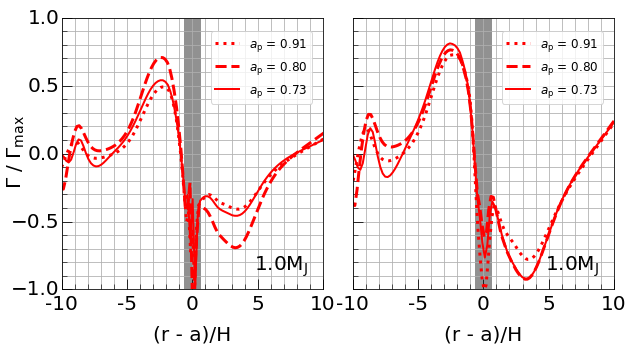}
\caption{Normalized torques for the same model as in Fig. \ref{fig7}, at the same times and
respective positions.
The left frame corresponds to non-accreting planets, while the right frame corresponds to accreting planets.
The gray area is the attenuated region within the Hill radius.}
\label{fig8}
\end{figure}

Figure \ref{fig8} shows the averaged azimuthal torques, which act on non-accreting (left panel) and accreting (right panel) planets, in their close vicinity. The torques were calculated for the standard parameter set. The magnitude of each torque is normalized with respect to the maximum value obtained along the radial distance. In both cases, the torques are stabilised in a few hundred orbits and reach the highest values in the vicinity of the planet (from -5 to +5 in the $x$--axis in Fig. \ref{fig8}). For both accreting and non-accreting planets, the plots shows that the torques  depend only slightly on the position of the planet in the disc.

Additionally, Fig. \ref{fig8} shows that the torque produced by the internal disc is proportionally stronger (of the order of 10\%) for accreting planets than for non-accreting planets. The similar behaviour was observed in the low viscosity regime. In the case of the high viscosity regime, we have noticed a positive torque which was closer to the accreting planets. The gray area in Fig. \ref{fig8} shows the region of attenuation defined by Eq. \ref{eqatenuacao}. For accreting planets, the torque within this region is smaller that is related to the decreasing mass due to the gas accretion by the planet. %For massive planets, the accretion of gas was high enough to decrease the torques in the vicinity of the planet \textbf{so that the maximum torque value  is outside the region shown in the plot.}

\begin{figure}
\centering
\includegraphics[width=1.0 \columnwidth,angle=0]{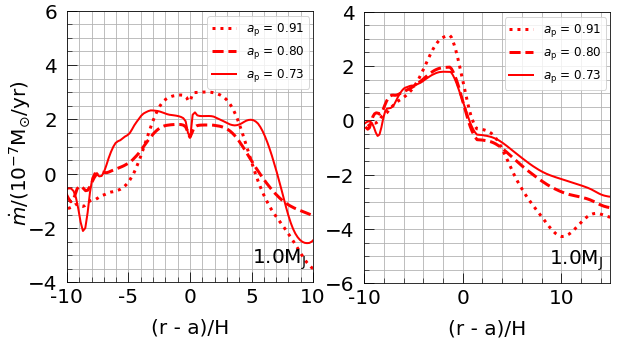}
\caption{Azimuthally averaged local accretion rate, $\dot{m}(r)$, at different positions of the planet
during the migration of the planet for the same model as in Fig. \ref{fig7}.}
\label{fig9}
\end{figure}

In Fig. \ref{fig9}, we show the mass accretion rate in the neighbourhood of the planet, $\dot{m}(r)$, at different positions of the migrating planet,  for non-accreting (left frame) and accreting (right frame) planets. For non-accreting planets (left panel), there is a large positive flow (contrary to the movement of the planet), which arises due to the fact that the migration velocity of the planet is higher than the viscous accretion speed. The same behaviour was observed in the case of low viscosity regime. On the other hand, in the high viscosity regime, with low accretion rates for the disk,  we have found some cases for which the gas flow is negative in the region close to the planetary gap. In this cases the migration velocity of the planet is lower that the viscous speed. These results are very similar to those obtained by \cite{Durman2015} for NIRVANA code, for non-accreting planets.

For accreting planets (Fig. \ref{fig9} right), we note a sharp fall of the gas flow in the accretion region. This is due to the fact that the planet is treated as a massive point in the hydrodynamic mesh, despite that the mass is withdrawn from the disc in the accretion region around the planet.
Analysing the mass flow around the planet in the case of high viscosity regime, we have noted that the fall of the gas flow is sharper due to the fact that the gas mass near the planet is smaller in this case. The observed asymmetry of the gas flow in the case of the accreting planet results in a distribution of gas around the planet, which is different from that around the non-accreting planet. As a consequence, the differential torque is smaller for accreting planets when compared to that for non-accreting planets.

Finally, in the case of low accretion rate of the disc and the high viscosity regime, the positive torque due to the internal disc is proportionally higher in regions close to the planet than in the case of the medium viscosity regime. However, the effect of attenuation of torques on the planet given by Eq. \ref{eqatenuacao} does not allow the reverse migration of accreting planets.

\subsection{Migrating planets with mass growth}\label{withgroth}

In this section we analyze the growth of a planet with the initial mass of $\simeq 1\mathrm{M_{J}}$. Since, in this case, we do not need to wait for the disc to attain the equilibrium state,  we allow the planet to start migrating after only 300 orbits, which is the time necessary to the disc stabilize in the planet's presence. In the first 300 orbits, the planet is fixed on circular orbit,  the gas is removed from the disc and the mass of the planet is kept constant.

To understand the influence of the planetary mass increase on the planet's migration process, we compare different scenarios of gas accretion onto the planet. In the first scenario, described in \cite{Kley1999}, the accretion rate is adjusted by the parameter $k_{a}$, which is a factor of increase or decrease of accretion. The second scenario corresponds to Russel's model \citep{Russell2011} described in Sect. \ref{modelaccretion}.
Two cases are observed: in case \textbf{R}, the mass is removed from the disc, but does not added to the planet, while in case \textbf{RA} the mass is removed and subsequently added to the planet. The linear moment of the accreted gas is not added to the planet, since its change does not interfere significantly  on the resulting process \citep{Durmann2017}. Finally, for the sake of comparison, we include a model  \textbf{N}, which does not consider for gas accretion, thus no mass is removed from the disc by the planet.

\begin{figure}
\centering
\includegraphics[width=\hsize]{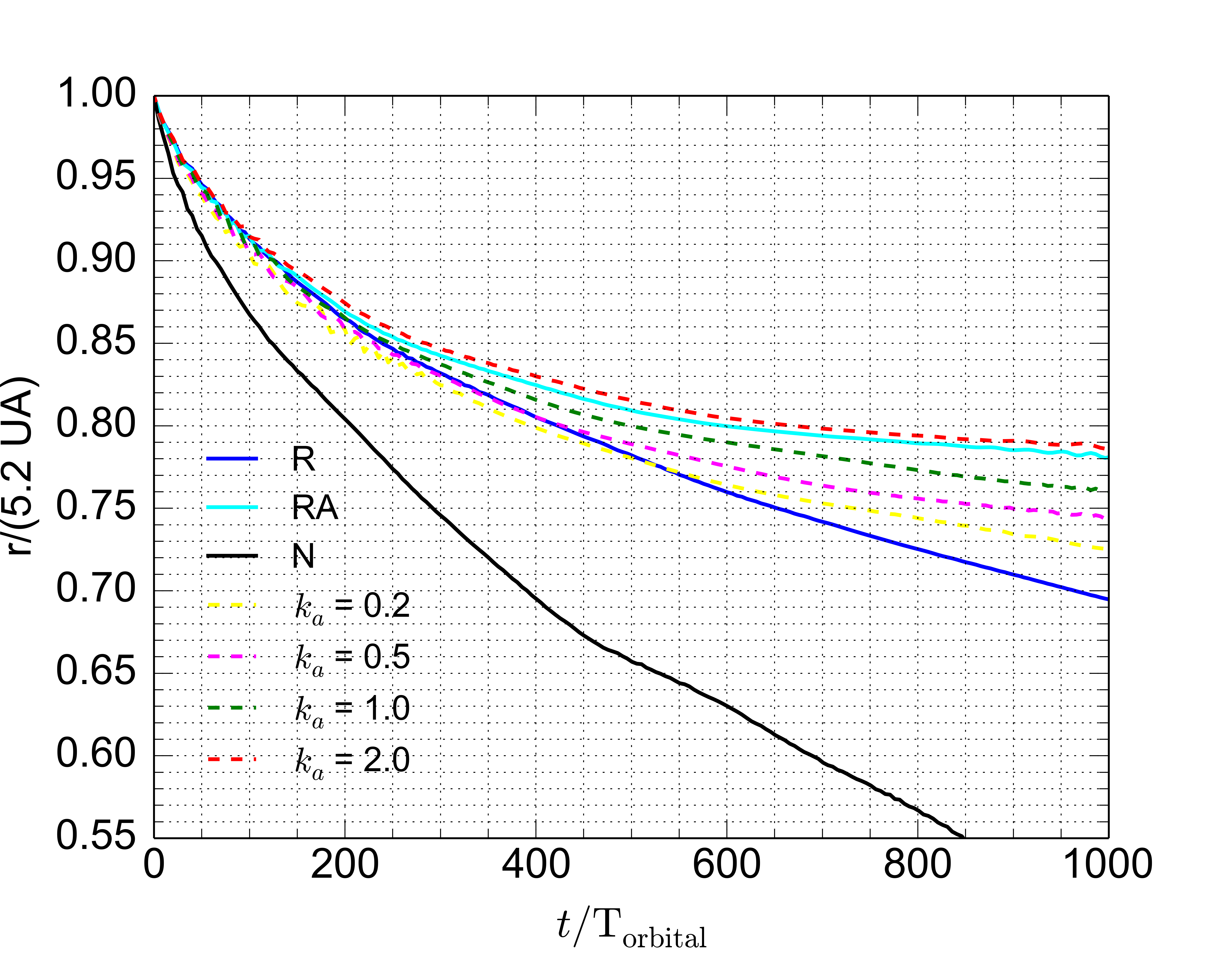}
\caption{Orbital decay of a planet with initial mass equal to $\simeq 1\mathrm{M_{J}}$. Model \textbf{N}: no matter
is removed from the gas disc. Model \textbf{R}: the gass is removed, without be added to the planet. Model \textbf{RA}:
the gas is removed according to Russell's approach (2011) and added to the planet. Kley's approach (1999)
is used with the different values of the parameter $k_{a}$, 0.2, 0.5, 1.0 and 2.0, when the mass is added to the planet.}
\label{fig10}
\end{figure}

The time evolution of the semimajor axis of the planet with initial mass equal to $\simeq 1\mathrm{M_{J}}$, in the different scenarios, is shown in Fig. \ref{fig10}. Comparing the \textbf{N} model (black curve) with the \textbf{R} (blue curve) and \textbf{RA} (cyan curve) models, we can observe that the removal of mass from the region around the planet decreases significantly the migration rate. Moreover, when the removed mass is added to the planet, the planetary inertia increases and, consequently, the the migration rate decreases. The decay of the semimajor axis in the model \textbf{RA} follows closely Kley's model with larger value of the parameter $k_{a}$ = 2.0. According to \cite{Durmann2017}, this value is a saturated upper limit of the accretion rate.

\begin{figure}
\centering
\includegraphics[width=\hsize]{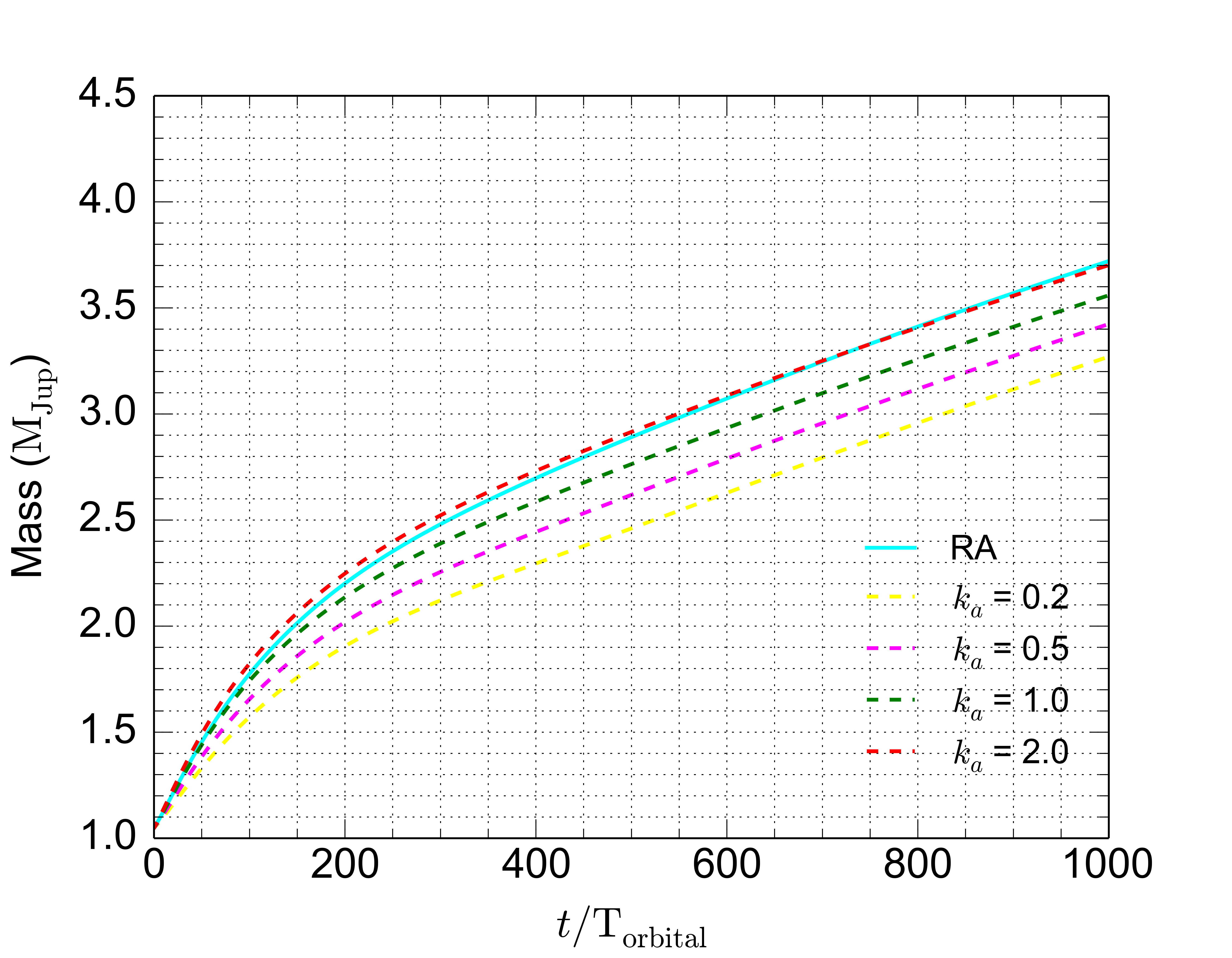}
\caption{Growth of a planet with initial mass equal to 1$\mathrm{M_{J}}$.
In the model RA the gas is removed according to Russell's approach (2011) and added to the planet.
In the model with $k_{a}=$ 0.2, 0.5, 1.0 and 2.0 the mass is removed using Kley's approach (1999)
and the mass too is added to the planet.}
\label{fig11}
\end{figure}

The growth of the planetary mass initially with $\simeq 1\mathrm{M_{J}}$ is shown in Fig. \ref{fig11}, where the continuous cyan curve corresponds to Russell's model \textbf{RA}, while the dashed curves correspond to Kley's model, with different values of the parameter $k_{a}$. We observe that, according to the model \textbf{RA},  the planet reaches the mass of more than 3.5$\mathrm{M_{J}}$ in 1000 orbits, which is very similar to what is observed using Kley's model with $k_{a} = 2.0$. This shows that, indeed, the runaway regime for gas accretion in \cite{Russell2011} is equivalent to the model in \cite{Kley1999} for high accretion rates.

\begin{figure}
\centering
\includegraphics[width=\hsize]{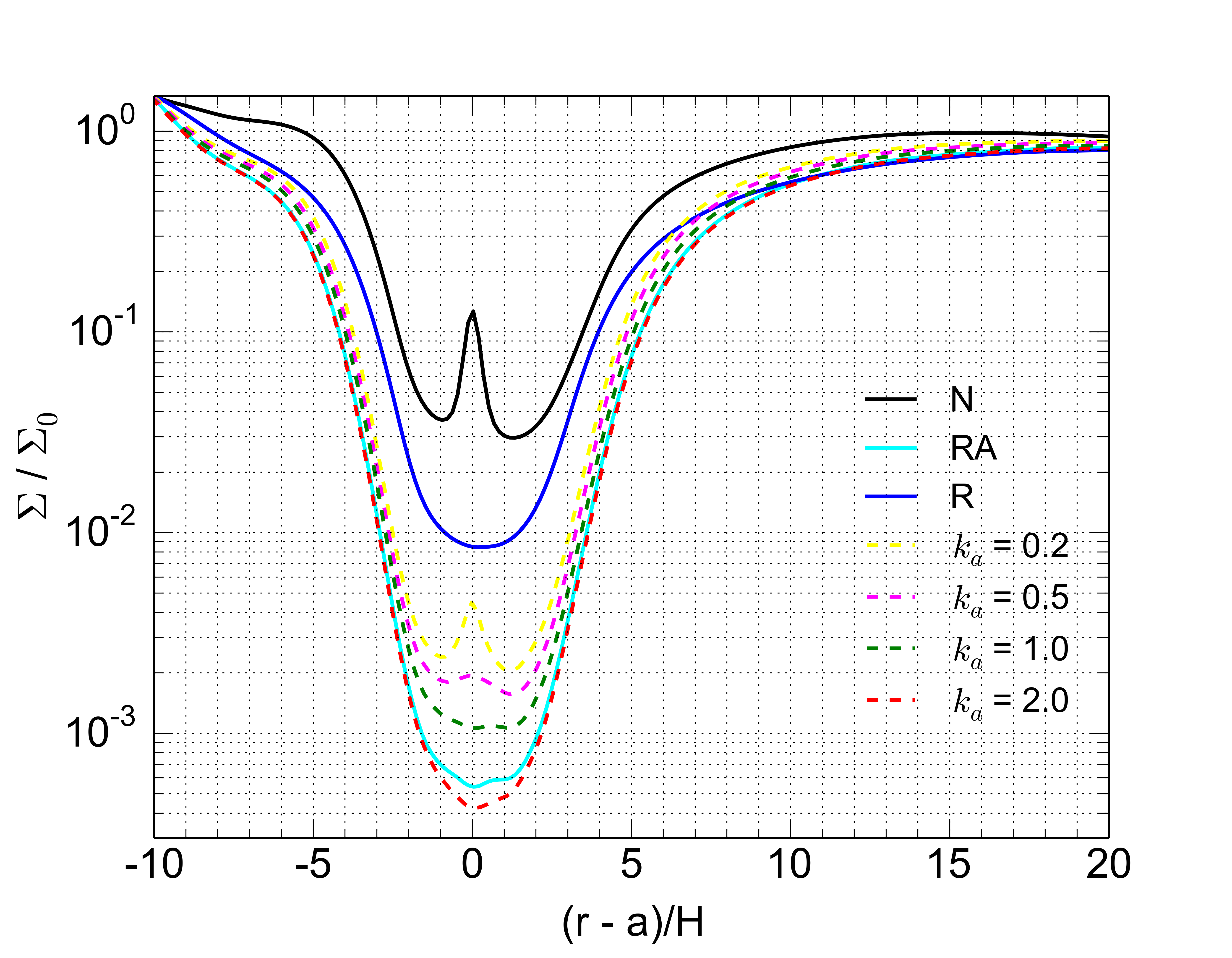}
\caption{Surface density profile close to the planet for the different models of accretion when the planet has
reached $r = 0.85R_{0}$.
Model \textbf{R}: the gass is removed, without be added to the planet. Model \textbf{RA}:
the gas is removed according to Russell's approach (2011) and added to the planet. Kley's approach (1999)
is used with the different values of the parameter $k_{a}$, 0.2, 0.5, 1.0 and 2.0, when the mass is added to the planet.}
\label{fig12}
\end{figure}

The surface density of the gas around the planet when the planet crosses $r = 0.85R_{0}$, is shown in Fig. \ref{fig12}, for the same models presented in Fig.\,\ref{fig10}. As expected, the gap is deeper when the planetary mass is higher at that instant, that happens for higher gas accretion rates. According to Fig.\,\ref{fig11}, the larger values of the mass correspond to the scenarios \textbf{RA} and $k_a$=2.0. Also, the higher accretion rates are responsible for the decreasing of the overdensity effect in the vicinity of the planet. Once again, we notice that the result obtained for the runaway regime with model \textbf{RA} is very similar to that obtained with $k_{a}$ = 2.0.

\begin{figure}
\centering
\includegraphics[width=\hsize]{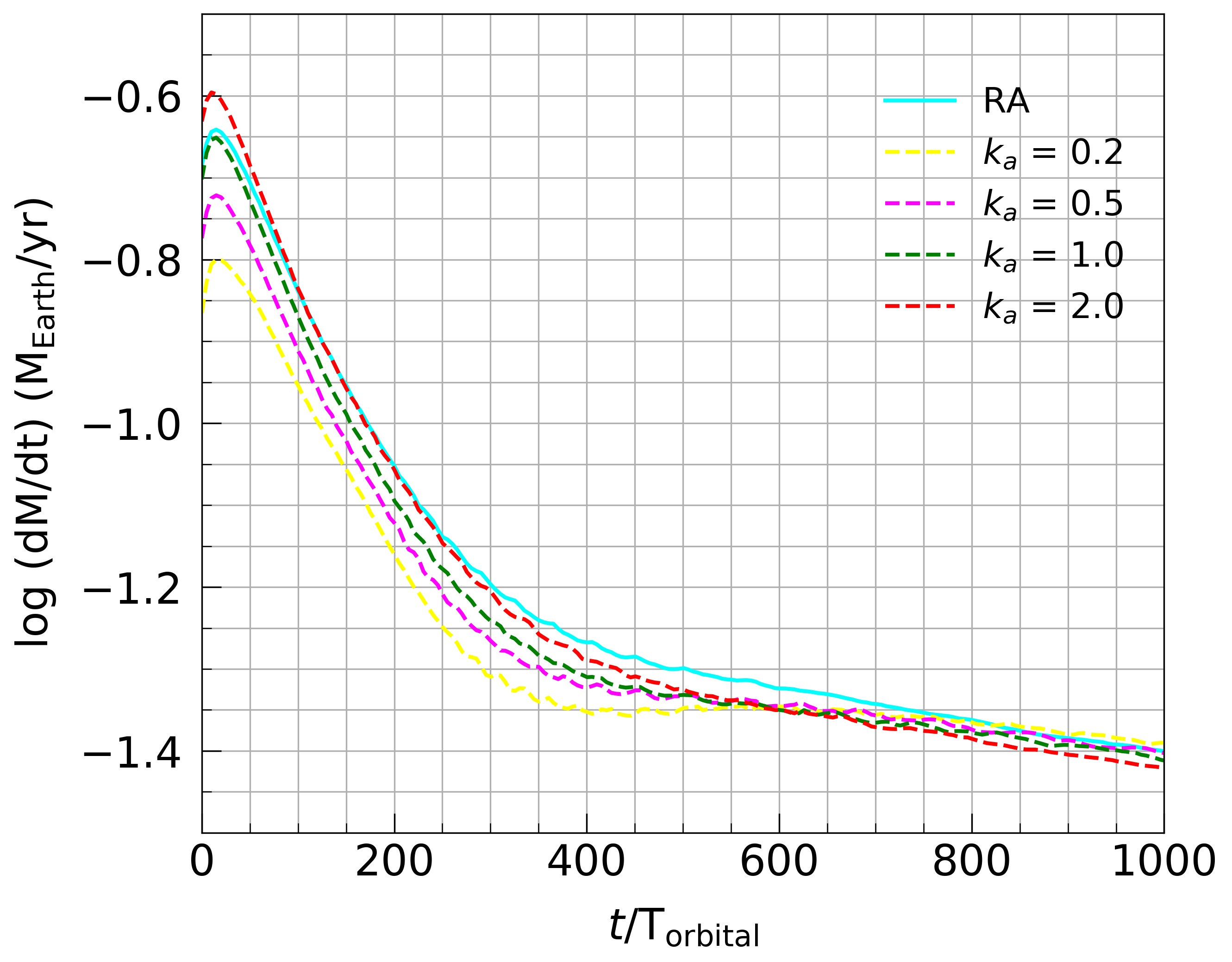}
\caption{Gas accretion rates as a function of time.
Model \textbf{RA}:
the gas is removed according to Russell's approach (2011) and added to the planet. Kley's approach (1999)
is used with the different values of the parameter $k_{a}$, 0.2, 0.5, 1.0 and 2.0, when the mass is added to the planet.}
\label{figtaxa1Mjup}
\end{figure}

Fig. \ref{figtaxa1Mjup} shows gas accretion rates for a growing giant planet. The behavior for the gas accretion rate in runaway regime is qualitatively very similar to the one obtained by \cite{Mordasini2012}, that uses a more complex model for gas accretion. 
However, in our model the planet can migrate and the disc model is different, so the results can not be compared in quantitative terms. A more detailed analysis that includes an adiabatic disk will be carried out in future works.

The results obtained show that Russell's model \citep{Russell2011} and Kley's model \citep{Kley1999} provide similar results when work with the giant planets in the runaway regime for gas accretion. However, the model in \cite{Kley1999} is not adapted to simulate the accretion of the low mass planets, because it does not account for  the Kelvin-Helmholtz scale. In contrast, Russell's model can be used to investigate the growth of the low mass planets since it adopts this approach.

\section{Growth of low mass planets}\label{lowplanets}

In order to study the accretion process in the case of low mass planets, we place a planet of mass $\simeq 20\mathrm{M_{Earth}}$ in a fixed orbit ($r$ = 1 code units) and analyze the increase of the planetary mass using the model for gas accretion described in section \ref{modelaccretion}.

\begin{figure}
\centering
\includegraphics[width=\hsize]{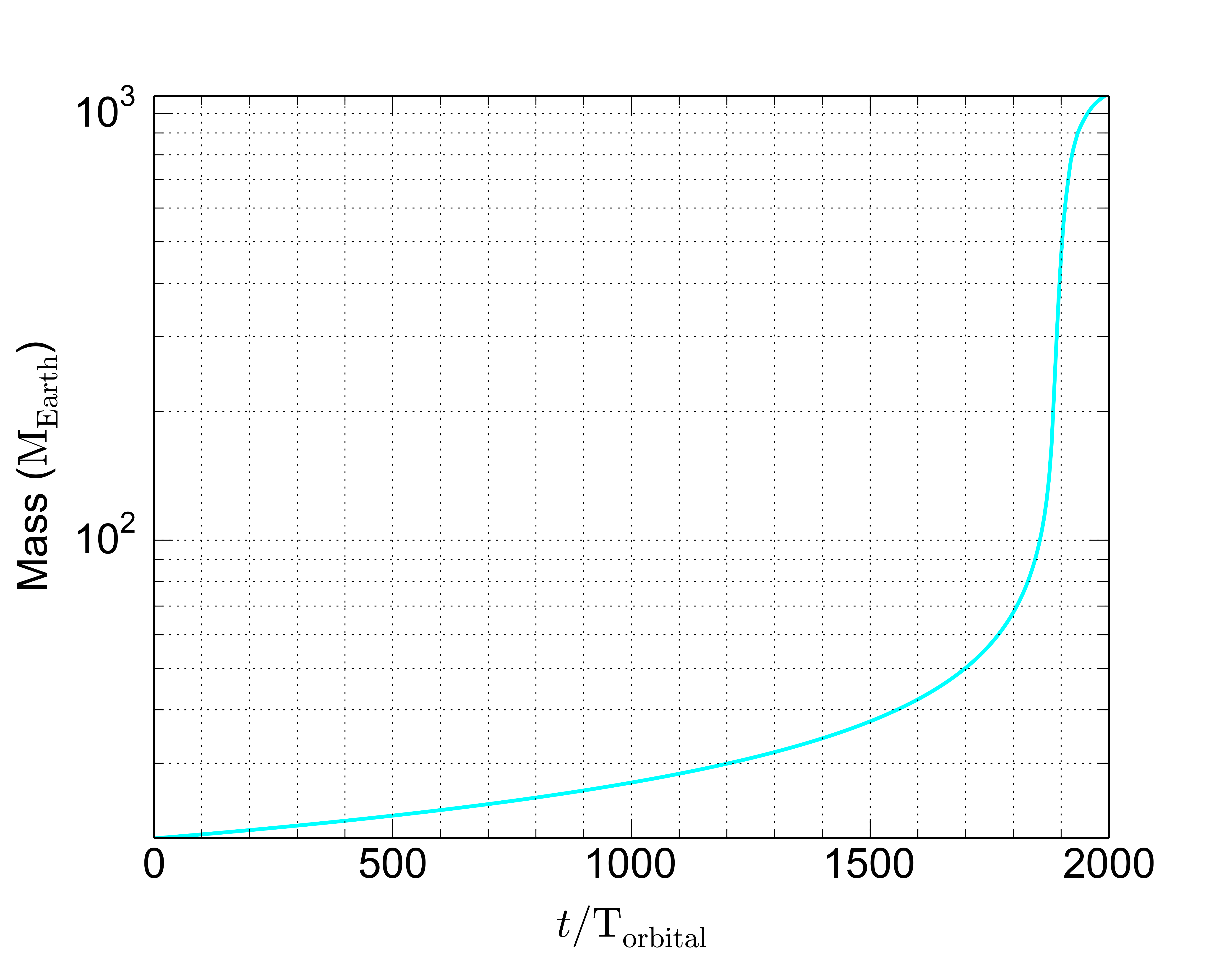}
\caption{In situ formation of initial mass planet $\simeq 20\mathrm{M_{Earth}}$ at $r$ = 1 code units using the
model for gas accretion described in section \ref{modelaccretion}.}
\label{fig13}
\end{figure}

Figure \ref{fig13} shows the in situ growth of the planet in the disc described by the standard parameter set from Table 1.  We observe that initially the planetary mass increases monotonously and slowly, but this process accelerates significantly when the planet enters in the runaway regime. This occurs  after approximately 1800 orbits, when the planetary mass reaches $\sim 0.2\mathrm{M_{J}}$, this behaviour is in agreement with other models for gas accretion \citep{Alibert2005,Fortier2013}. It should be noted that any model which considers only runaway accretion regime, would not be able to reproduce this behaviour, since, in this case, the formation time would be  very short. The runaway accretion decreases the time for the planet to establish type II migration, since the gas along the planetary orbit is quickly exhausted.

\begin{figure}
\centering
\includegraphics[width=\hsize]{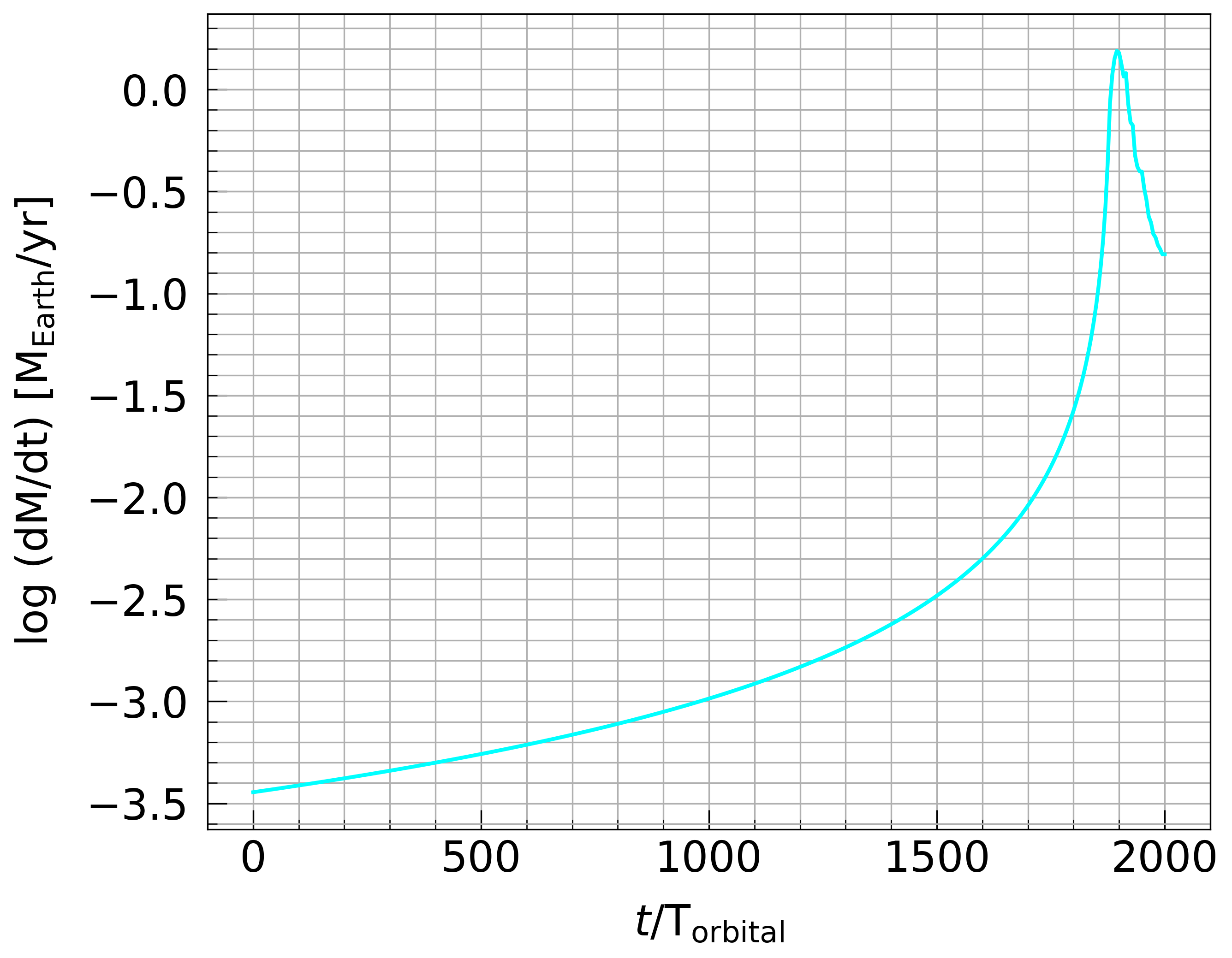}
\caption{Gas accretion rates as a function of time for
the situ formation of initial mass planet $\simeq 20\mathrm{M_{Earth}}$ at $r$ = 1 code units 
using the model for gas accretion described in section \ref{modelaccretion}.}
\label{taxa20MT}
\end{figure}

Fig. \ref{taxa20MT} shows gas accretion rates for the planet fixed in $r$ = 1 code units. 
The behavior for the gas accretion rate is similar to the one obtained by \cite{Mordasini2012}, 
that uses a more complex model for gas accretion. The only difference observed is the scale of the accretion rate, which is 
due to the fact that our model for the gas disk has a much higher surface density, thus resulting in higher accretion rates. 
In \cite{Mordasini2012} it is also possible to find a complete scenario that 
includes the formation of the solid core. Here, we already start with a mass of $\simeq 20\mathrm{M_{Earth}}$. 
A complete scenario including a solid accretion model within 
the FARGO3D is under development and will be analyzed in future work.

\begin{figure}
\centering
\includegraphics[width=0.76 \columnwidth,angle=0]{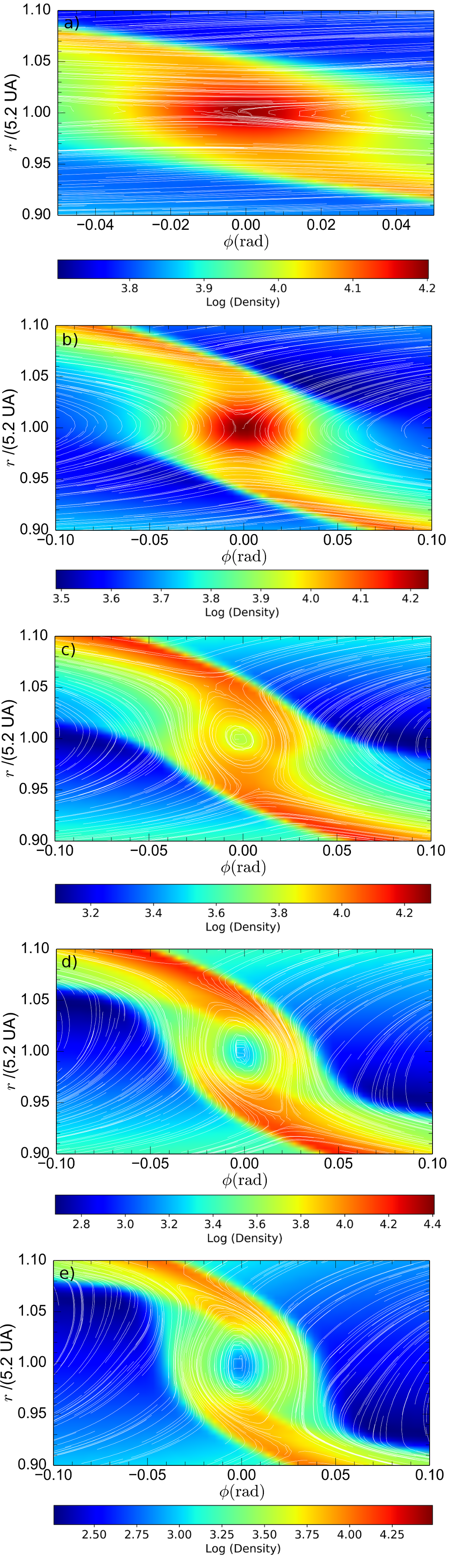}
\caption{The grid mesh colored with the log of the superficial density (in $\mathrm{kg/m^{2}}$) and with
the streamlines in the planet's frame. (A) $\simeq 30\mathrm{M_{Earth}}$
($t$ = 1210 orbits), (B) $\simeq 0.2\mathrm{M_{J}}$ ($t$ = 1800 orbits), (C) $\simeq 0.5\mathrm{M_{J}}$ ($t$ = 1880 orbits),
(D) $\simeq 1.2\mathrm{M_{J}}$ ($t$ = 1895 orbits) and (E) $\simeq 2.0\mathrm{M_{J}}$ ($t$ = 1910 orbits).}
\label{fig14}
\end{figure}

We analyze the effect of gas accretion on the movement of the gas around the planet by performing a streamline analysis, shown in Fig. \ref{fig14}. In Fig. \ref{fig14}a, we observe that, at the beginning of the accretion process, the accreted gas comes from regions near the planet. Moreover, we notice a high gas density region around the planet. In Fig. \ref{fig14}b, the gravitational perturbations due to the planet tend to clear the gas along the planetary orbit, at the same time increasing the gas density in the region close to the planet. The accretion rate of gas is still too slow to decrease the gas density close to the planet. As the planetary mass increases, the gas mass tends to accumulate in the spiral arms. The mass transfer from the outer disc to the inner disc in the region close to the planet is still small.

In Fig. \ref{fig14}c, the planet starts to accrete the gas in the runaway regime; we see the mass depletion in the region around the planet, and the gas density around the planet tends to decrease. Following, the planet begins to accrete gas from the connection of the inner disc with the outer disc. In Fig. \ref{fig14}d, we see a stronger mass transfer between the inner and outer disc. The gas tends to surround the planet and the high rate of gas accretion tends to clear the nearby region of the planet. Finally, in Fig. \ref{fig14}e, an amount of gas surrounds the planet, where it is accreted to the planets, while the rest of the gas is accelerated by the planet and driven to the inner disc. Thus, we can observe an intense mass flow between the inner and outer discs.

In summary, much of the mass during the early stages of accretion comes from regions along the orbit of the planet. However, with the increase of the planetary mass and the clearing of the orbital region, a connection between the inner and outer discs is established, from where the mass is accreted now. In any case, despite the accretion of gas, the time needed for the planet to clear its orbit is still very long. Thus, type I migration would be sufficient to drive the planets close to the star until the planet enters a type II migratory process.

In the model investigated in the paper, the gas accretion process for low-mass planets does not depend substantially on the position of the planet in the disc. It should be stressed that we do not consider solid core growth due to accretion of planetesimals that is a process which is strongly dependent on the density of solids along the disc. Therefore, it is necessary to consider accretion of solids to increase the rate of mass accretion. In addition, the accretion rate of solids is important in determining the critical mass for which gas accretion begins \citep{Ikoma2000}. A model for accretion of solids will be elaborated in future work.

\section{Conclusions}

In this work we studied the migration of massive planets in locally isothermal disc in equilibrium using the FARGO3D code. Initially, the position of the planet was fixed and the structure of the gap was analyzed 
for different values of the planetary mass and disc viscosity (Fig. \ref{fig2}). We obtained that the depth of the gap increases with planet mass and reduces with increasing viscosity. This result is in agreement 
with that obtained by \cite{Durman2015} using the NIRVANA code for non-accreting planets. For accreting planets, the gaps are deeper, while the overdensity regions close to the planets are vanishing. 
The results above are independent of the initial accretion rate of the disc.

Analyzing the flux through the gap during the migration process, we observe rapid inward migration in the case when the planet decays faster than the typical viscous inward drift of the disc material. We find that, 
for smaller disc mass and higher viscosity, the planet decays more slowly than the typical viscous inward drift of the disc material. 

In this paper, we performed simulations for both non-accreting planets and accreting planets. In latter case, we adapt a model for gas accretion from \cite{Russell2011}, to calculate the rate of accretion. Our main goal is to use a more rigorous and robust gas accretion scheme that is valid and consistent for a larger range of planetary masses. In this model, the size of the accretion zone is obtained taking into account the planet's Bondi radius and the Hill radius; in this way, physical characteristics of the disc, 
such as its thermal velocity, become important to determine which cells belong to the accretion region. The timescales involved in accreting gas onto a planet are also calculated cell-by-cell using the orbital period about the planet instead of the predefined orbital period in the predecessor code. This modification allows us to obtain an accretion rate in the runaway regime that is closer to the Bondi accretion limit. 

Also, we take into account the Kelvin-Helmholtz collapse time-scale in approximation of the planet's gas accretion rate for small masses. This modification allows us to study the gas accretion process for planets which have not yet reached the runaway regime.

As expected, the reduced mass in the vicinity of the planet reduces migration rate for accreting planets. The difference in the values of the migration rates becomes less significant with the increase of the planetary mass (Fig. \ref{fig5}, right) and with the decrease of disc accretion rate (Fig. \ref{fig5}, center).
Also, the effect of the gas accretion is more significant in higher viscosity regimes (Fig. \ref{fig5}, left). In fact, in low viscous discs, the mass tends to accumulate faster in the regions around the planet and accretion can not account for cleaning the region in order to  change significantly the differential torque. Minor planetary masses are more affected from the accretion process, because the smaller the mass of the planet, the smaller the gravitational effect that prevents the mass from entering the gap, therefore, the process of accretion tends to clean the region of the planet favoring the appearance of the planetary gap.

% Low mass discs can arise at the end of the planetary formation phase due
% to gas accretion and photoevaporation processes. In this case, it is possible that
% the planetary migration rate lower than evolution viscous of disc.

In this work, we study only 2D discs. For massive planets, the averaged gap
profile is identical for 2D and 3D discs and we may expect very
similar effects for the planetary migration. Nevertheless, it may be interesting to perform
3D simulations and analyze the migration properties because little is known yet 
about the mechanism of gas accretion in the 3D case.

The size of the accretion zone depends on the thermal velocity of the gas. 
Recent works show the importance of considering the thermodynamical effects involved in the gas accretion process. Changes in entropy can change the luminosity and the geometry of the accretion zone, thus modifying the gas accretion rate of the planet. Recent 3D hydrodynamic simulations show that supersonic shock fronts are important for analyzing the thermodynamics of the accretion process \citep{Marley2007,Szulagyi2017}. 
Thus, in future works, it would be interesting to explore the changes
that radiative discs could provoke in the accretion zone. Models with larger numerical
resolutions would be needed in this case.

We test the model of gas accretion for the runaway regime by considering a 1$\mathrm{M_{J}}$ mass planet migrating
towards the star and accreting mass during its motion. We note that the runaway regime
corresponds to that used by \cite{Kley1999} for parameter $k_{a}$ = 2.0. This agrees
with the expected result given that the accretion of gas is close to the saturated regime for this parameter.

Also, we simulated the growth of a low mass planet fixed at position $r$ = 1 code units and we evaluated
the time to enter the runaway regime and open the gap needed for type II migration. We found a time of
approximately 1800 orbits, which is very high considering the rapid migration of type I.
Thus, the planet would fall in the star before being able to open the gap. However, this has to be
yet further investigated. In fact, new parameters for the gas disc may reveal a faster gap
opening process, specially on heat transfer discs. In addition, a complete model that includes the
accretion of planetesimals could give us an indication of when the gas accretion process begins.
These factors will be explored in future work.

\section*{Acknowledgements}

The authors are grateful to P. Ben\'itez-Llambay for meaningful discussions and suggestions.
This work was supported by FAPESP (Brazil) through the grants 2014/00492-3.
This work has used the computing facilities of the Laboratory of Astroinformatics
(IAG/USP, NAT/Unicsul), that were purchased thanks to the Brazilian
agency FAPESP (grant 2009/54006-4) and the INCT-A.

%%%%%%%%%%%%%%%%%%%%%%%%%%%%%%%%%%%%%%%%%%%%%%%%%%



%\bibliographystyle{mnras}

% \bibliographystyle{mnras}
% \bibliography{bibliografia.bib}

%\begin{thebibliography}{99}
%\end{thebibliography}

%%%%%%%%%%%%%%%%%%%%%%%%%%%%%%%%%%%%%%%%%%%%%%%%%%

% Don't change these lines
\bsp	% typesetting comment
\label{lastpage}
\end{document}